# A Digital Engineering Approach to Testing Modern AI and Complex Systems

Joseph R. Guerci[1], *Fellow, IEEE*, Sandeep Gogineni[1], *Senior Member, IEEE*, Robert W. Schutz[1], Gavin I. McGee[1], Brian C. Watson[1], Hoan K. Nguyen[1], *Senior Member, IEEE,* John Don Carlos[1], Daniel L. Stevens[2,] *Senior Member, IEEE*

***Abstract***— Modern AI (i.e., Deep Learning and its variants) is here to stay. However, its enigmatic "black box" nature presents a fundamental challenge to the traditional methods of test and validation (T&E). Or does it? In this paper we introduce a Digital Engineering (DE) approach to T&E (DE-T&E), combined with generative AI, that can achieve requisite mil-spec statistical validation as well as uncover potential deleterious "Black Swan" events that might otherwise not be uncovered—until it's too late. An illustration of these concepts is presented for an advanced modern radar example employing deep learning AI.

## I. INTRODUCTION: STATISTICAL NATURE OF DOD TESTING AND MIL-SPEC

DoD testing and Mil-Spec have long since recognized the stochastic nature of modern weapon systems [1-3]. Thus, achieving acceptable performance has always been couched in statistics. Take for example weapon system operational availability dubbed "A-sub-o", $A_o$ [4]. It refers to the percentage of time a weapon system is ready for use when called upon. For example, weapon system X shall achieve and $A_o$ of 97% with 95% confidence. Thus, at its core, mil-spec does treat a system under test (SUT) as a "black box". Statistical performance and reliability in a representative operating environment are what matter.

Ironically, the amount of testing required in a representative warfighting environment to achieve such confidence practically precludes actual sample data for all but the simplest of components. Thus, a combination of partial field test data, combined with analytical and other empirical data is employed as a substitute for "real" test data. Clearly, less than ideal. However, the "black box" nature of deep learning AI precludes this *ad hoc*, quasi analytical approach. One cannot decompose a CNN using traditional constructive systems engineering methods (e.g., an interconnected series of definitized functional blocks). As we shall see in the next section, modern DE tools can provide an approach that solves this issue for a broad class of AI systems, as well as other highly complex (not necessarily AI) systems. This will lead to better T&E methods for all systems.

## II. A DIGITAL ENGINEERING APPROACH TO TESTING & VALIDATING AI AND OTHER COMPLEX SYSTEMS

Modern Digital Engineering (DE) [5] is a culmination of decades of advances in model-based systems engineering (MBSE) and advances in high performance computing (HPC). The latter enables ever finer physics-based, high-fidelity models that can produce useful results in an acceptable period of time. Given the decades in time, resources and manpower devoted to advanced aeronautical systems, it is no surprise that projects such as the Next Generation Air Dominance (NGAD) fighter and B-21 bomber have led the way [6-7]. Indeed, the Office of the Secretary of Defense (OSD) has recently issued a department wide directive "Instruction 5000.97" mandating that all new programs "Will incorporate Digital Engineering…" [8]. While the DE tools for aerodynamics and stealth have been refined for decades and thus can be relied upon to produce accurate modeling results, the same cannot be said in general for the avionics systems carried by such platforms (there are exceptions as discussed below).

While having an accurate "Digital Twin" for a "system under test" is an essential prerequisite for a DE approach to T&E, to provide the requisite sample statistics to achieve a given confidence, an accurate digital twin for the operating environment is also required. Thus, the prerequisite elements of a DE approach to T&E consist of the following:

- A digital twin for the system under test. In the case of an aircraft for example, this would include both the airframe and avionics.
- A digital twin for the environment. In the case of a fighter aircraft, for example, this would include atmospherics, terrain, and the environmentals associated with any and all sensor/comms systems (radio frequency (RF), electro-optical/infrared (EO/IR), etc.)
- An estimate of an adversary's order-of-battle and capabilities. Again, for a fighter aircraft, this would include models of the integrated air defense systems (IADS). Tools such as AFSIM are examples of this type of model [9].

[1]Information Systems Laboratories, Inc., San Diego, CA USA
[2]Air Force Research Laboratory, Information Directorate, Rome, NY USA

In the next section, we outline a new end-to-end method for achieving the above.

## III. A NEW DIGITAL ENGINEERING METHOD FOR T&E

In this section, a new three phased approach to T&E of AI and complex systems is developed. In Section 3.2, this approach is applied to an advanced RF application employing deep-learning AI. The three phases are: (I) Baseline; (II) Excursion; and (III) "Black Swan".

*Phase I Baseline:*
In this Phase, the baseline digital twin models for the system under test (SUT) and the environmentals are assumed to be accurate (modeling errors are addressed in Phase II). Extensive Monte Carlo (MC) sortie analysis is conducted to achieve statistical convergence. To be practical, requisite HPC resources are also assumed to be available. These results are combined with whatever "real" data is available, along with other analyses, to achieve a baseline performance metric. If the desired performance (e.g., $A_o$) is achieved, the process proceeds to Phase II. If not, designers will need to determine what modifications need to be made based on the performance results.

*Phase II Excursion:*
No models are perfect. In this phase, excursions from baseline models are conducted to ensure robustness of the results. Since excursions represent, by definition, low probability events, there is no need to completely recreate all of the MC sorties in Phase I. This phase could also be characterized euphemistically as the "known unknowns" phase. While not likely, these excursions are certainly possible given what is known and unknown. For example, a jammer's effective radiated power (ERP) could be higher than assumed. What's the impact? After the SUT successfully passes Phase II, success is declared. So, what is the need for a Phase III? There can also be "unknown unknowns". This is addressed in Phase III.

More specifically in Phase II, let $\Omega_R$ denote the nominal set of training data used in Phase 1, with $\|\Omega_R\|_0 = N_R$, the number of MC runs to achieve convergence, and $\|\Omega_R\|_{div} = D_R$ denote the "diversity metric" of the nominal data set. This is a generalization of the usual variance concepts. An increase in diversity and quantity are performed in Phase II to access the robustness of the baseline design SUT. Specifically, $\kappa_1, \kappa_2$ are chosen such that

$$\|\Omega_S\|_0 = \kappa_1 \|\Omega_R\|_0 = \kappa_1 N_R \quad (1)$$
$$\|\Omega_S\|_{div} = \kappa_2 \|\Omega_R\|_{div} = \kappa_2 D_R \quad (2)$$

where $\kappa_1, \kappa_2 > 1$, and $\|\Omega_S\|_0, \|\Omega_S\|_{div}$ denote the new augmented data sets.

In words, additional synthetic data is produced with excursions from the nominal baseline model and additional MC runs are performed until performance convergence is again achieved. An assessment is then performed to determine if further design iteration is required to provide robustness against unknown modeling errors. These concepts are made more concrete with an example in Section 4.

*Phase III "Black Swan":*
The term "Black Swan" refers to events that are so far removed from the norm, a priori, that they were not even thought of (much less thought possible), but which can nonetheless have devastating consequences [10]. They are also events which, in hindsight, are realized to indeed have been possible. Basically, by definition, they are not predictable by humans—hence the moniker "unknown unknowns". With the advent of generative AI (GAI) [11], it is now possible to include an entirely new phase into the T&E process: A Black Swan search stage. The idea is to use GAI to create scenarios which, while constrained by the known laws of physics and information theory, are nonetheless far from the "norm". This Phase III is not meant to be an additional gate that a SUT must pass to achieve acceptance, rather it is meant to be a phase run concurrently with deployment. If a significantly deleterious event is discovered, designers can decide what if any corrective action needs to be taken proactively. Further details of this Black Swan stage are presented in Section 4.

In the next section, we illustrate the above with a relevant radio frequency (RF) example employing advanced deep learning AI.

## IV. AN EXAMPLE OF APPLYING THE NEW DE-T&E APPROACH

Given the above new DE T&E theoretical framework, we have chosen a radar application to demonstrate this method for two reasons: (1) The availability of advanced RF DE tools (e.g., RFView [12], see Figure 1); and (2) Relevancy to modern operations. More specifically, the goal is to localize targets in the range-Doppler space in a scenario corrupted by complex ground clutter. Multiple deep-learning convolutional neural networks (CNNs) were trained to take as inputs the range-Doppler plots generated by a ground moving target indicator (GMTI) radar and produce an output that would be the estimated range and Doppler corresponding to the target [13]. With RFView, digital twins can be developed for the baseline radar, i.e. the system under test, and the propagation/clutter environment. These, in turn, can be used to validate the deep learning AI detection algorithm.

To train the CNNs, we leveraged RFView which served as the Digital Twin for the SUT as well as the RF environment. RFView was used to generate several realizations of range-Doppler maps with realistic site-specific clutter representations. The target of interest was randomly moved across the range-Doppler space to ensure the CNN sees sufficient diversity in the dataset. The CNN trained on this dataset was then evaluated

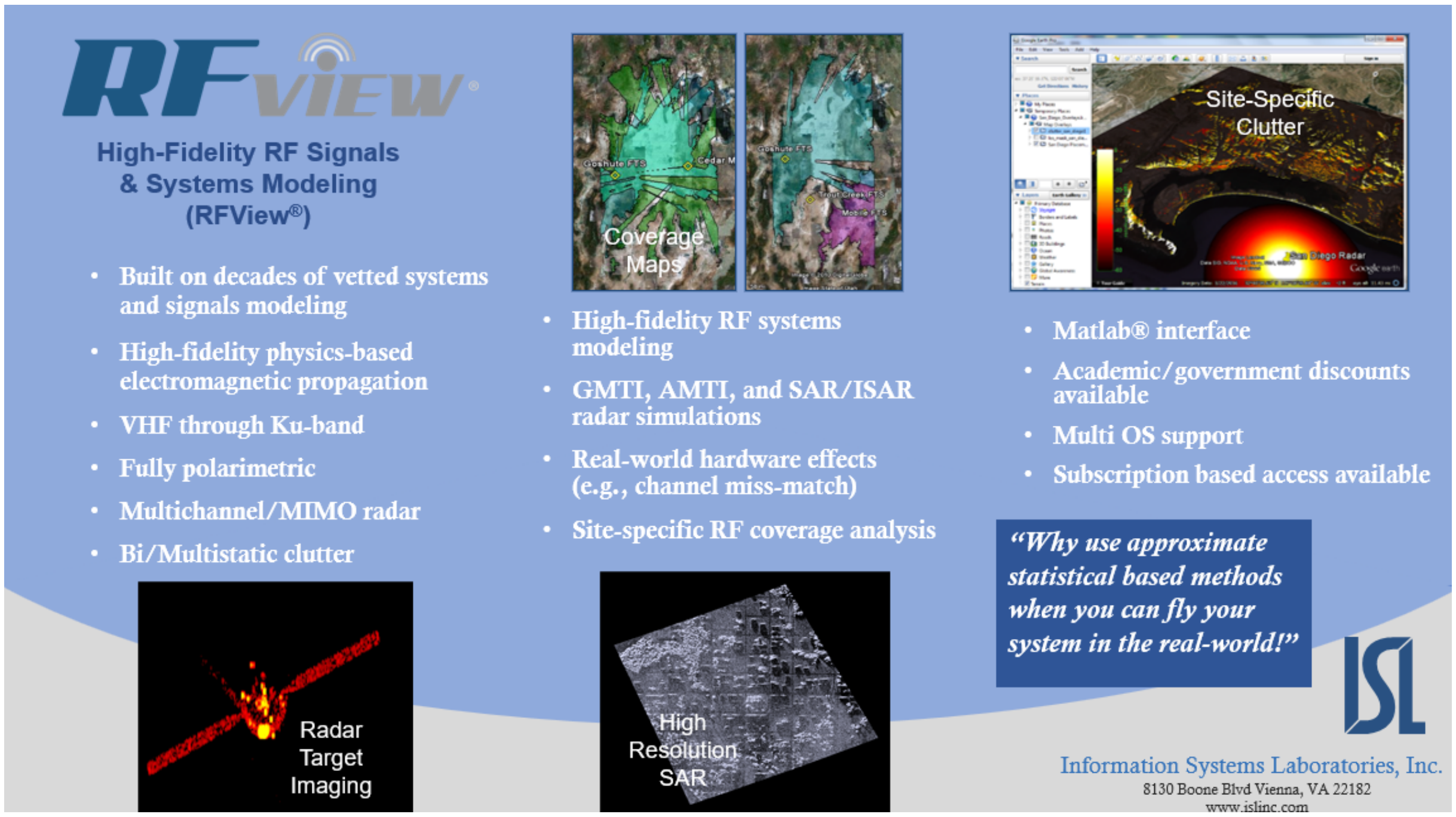

Fig. 1. RFView Capabilities

using the steps of the new DE T&E process described above.

In Phase I of the new DE process, a large validation dataset was generated using RFView to perform the necessary MC analysis. Upon convergence, performance analysis of the CNN was conducted. System design variables such as transmit power, antenna size, etc., were adjusted until the performance met the prescribed goal, which in this case was the estimation error in both the range and Doppler dimensions. Since no model is perfect, to perform the sensitivity analysis in Phase II, we introduced modeling errors in the clutter generation engine of RFView by perturbing the scattered power from individual clutter patches in the scene. This increased the clutter signal strength and further obscured the target of interest, thereby resulting in performance degradation. To perform this step, a second dataset was generated which was primarily used for evaluation of the CNN under these model excursions.

The level of performance dropped as a result of these excursions was then captured and necessary measures to offset, such as increasing the antenna size, were performed. By increasing the antenna size, the radar beam becomes more focused and hence results in better suppression of the increased mainbeam clutter response [13]. At this stage, since the system variables have been changed, we regenerated new testing and validation datasets for the desired CNN. This is the third dataset as part of the overall DE-T&E process. This re-trained CNN was then tested to ensure it meets the prescribed performance criteria. This new process ensures that the CNN not only meets the desired estimation accuracy for the original digital twin, but that it also achieves performance thresholds even when the digital twin introduces additional modeling errors. In this step, we ensured that the re-designed CNN was tested on clutter data with and without the modeling perturbations.

A key to demonstrating the process above is the generation of high-fidelity physics-based datasets for the GMTI application in realistic ground clutter using RFView. For the baseline scenario, we simulated an airborne radar flying along the coast of southern California. The parameters of the radar platform are presented below:

Latitude: 32.4005° N

Longitude: 117.1993° W

Height: 1000 m AGL

Speed: 100 m/s

The platform is flying due north and parallel to the earth's surface. The clutter scene that is being simulated is centered at a latitude of 32.5505° N and a longitude of 117.0493° W. Note that the target that is being simulated is moving on the ground and the target is not necessarily located at the center of the scene. For each realization of the dataset, the target location latitude was chosen between 32.5439° N and 32.5571° N while the longitude was chosen between 116.9577° W and 117.1406° W. In addition to the location, the ground moving target speeds were either 7 m/s or 14 m/s. Note that even though the target speeds were one of two values, the target does induce several possible Doppler frequency shifts in the range-Doppler (RD) map based on its location during that sample. The random location and speed of the target results in randomness of both the range and Doppler within the RD map. The dataset consisted of 5000 RD maps along with their labels extracted from RFView.

wavelength). The elemental data was coherently combined based on the look angle for each sample.

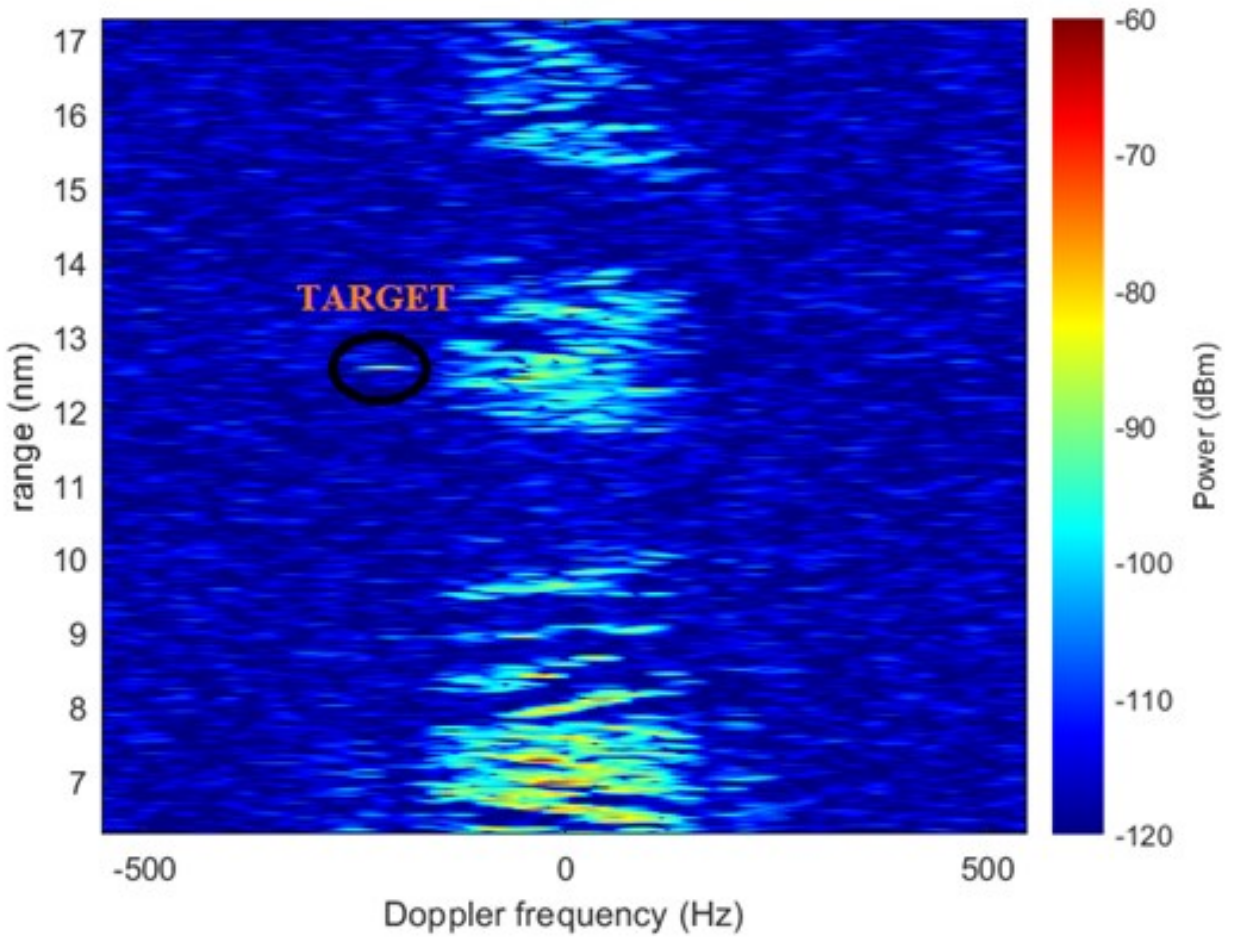

Fig. 2. Range-Doppler plot example for GMTI application. In this case, the target is well separated from the mainbeam clutter returns.

| Layer | Output Size | Parameters |
|---|---|---|
| Input | 680x320x1 | 0 |
| Convolutional | 340x160x8 | 112 |
| Convolutional | 170x80x32 | 2656 |
| Convolutional | 85x40x64 | 4992 |
| Convolutional | 43x20x128 | 9408 |
| Convolutional | 43x20x256 | 18176 |
| Convolutional | 22x10x256 | 35200 |
| Convolutional | 11x5x256 | 69120 |
| Convolutional | 6x3x256 | 69120 |
| Convolutional | 3x2x512 | 135936 |
| Dense | 256 | 786688 |
| Dense | 2 | 514 |

Fig. 4. MobileNet architecture used for GMTI target localization.

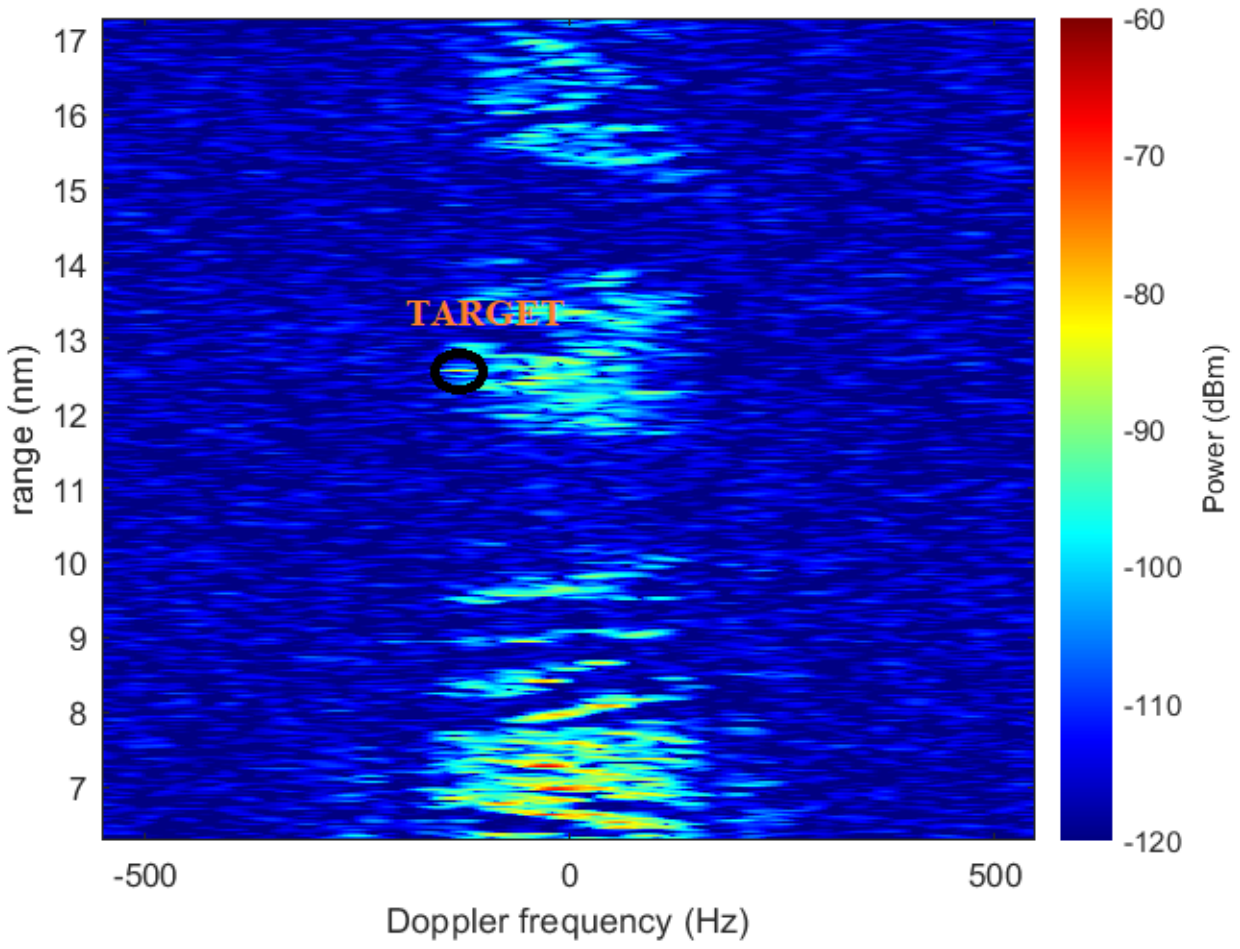

Fig. 3. Example where target is obscured by mainbeam clutter.

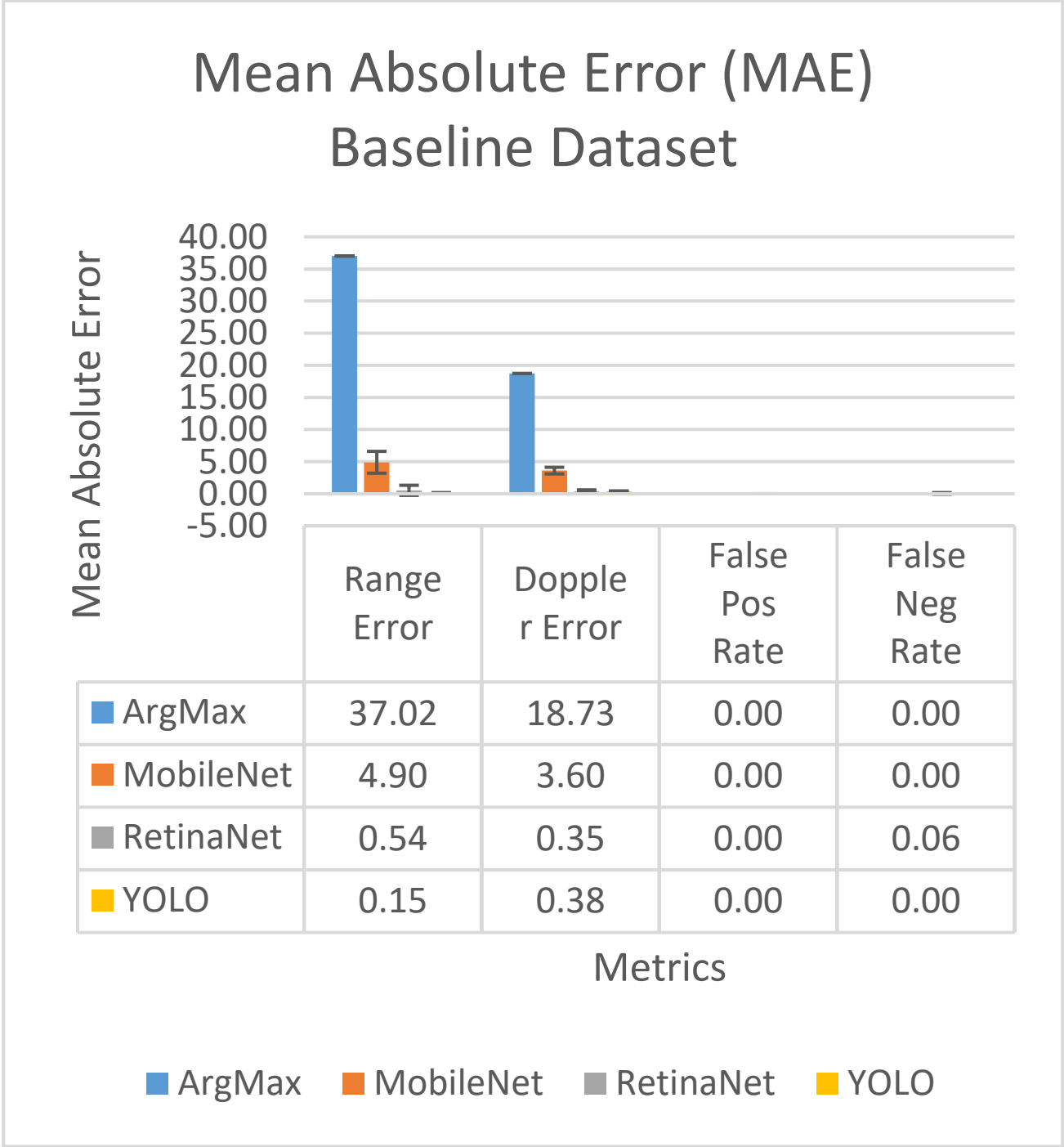

Fig. 5. Mean Absolute Error (MAE) for different algorithms along with the corresponding 1-sigma error bars

In Figure 2, we plot a sample RD map with clutter and target of interest. Here, we observe that the high-fidelity simulated data has site-specific clutter returns. Because the platform itself is moving, clutter returns also undergo a Doppler shift [13]. The target, which is highlighted with a black circle in the image below, can be distinguished from the background clutter in this example due to its larger Doppler shift relative to ground clutter. In general, however, it depends on the target RCS as well as the strength of the background clutter, speed of the target, etc. In Figure 2, the target was moving at a speed of 5 m/s. When the target is changed to a slow moving one at a velocity of 3 m/s, we observe from Figure 3 that the target is now no longer straight forward to distinguish from the background clutter. The examples in Figs. 2 and 3 are just for illustration purposes. The complete dataset included clutter returns for various antenna look angles and target locations. For the generated baseline dataset, the antenna array that was simulated has 10 horizontal elements and 5 vertical elements. Interelement spacing for this X-band (10GHz) simulation was the maximum of 1.5 cm to avoid grating lobes (half the

**Phase I**

Again, in Phase I of the DE process, the best estimate of all digital twins (SUT, environment, adversarial actions, etc.) is used to determine system efficacy, and baseline weapon system availability $A_o$. Before discussing the Phase II excursion results, we will first provide further details on the training of the AI CNNs used as stand ins for target detection and localization. We used three different CNNs to train on the aforementioned dataset for identifying the target location from RD maps that have clutter returns.

The first network was a version of MobileNet described in Figure 4. The baseline dataset above was used to train a 12-layer CNN to localize the target of interest within a Range-Doppler plot. The architecture of the MobileNet includes several convolutional layers which also include a rectified linear unit (ReLU) activation, a batch normalization, and max pooling [14]. Figure 4 delineates the MobileNet architecture used in this project. RetinaNet and YOLOv7 were also used to detect the targets of interest from the RD maps. Note that the three phased approach presented in this paper is independent of the specific algorithms. We have chosen these three well-known CNN algorithms for object detection for demonstrating the different stages of the T&E process.

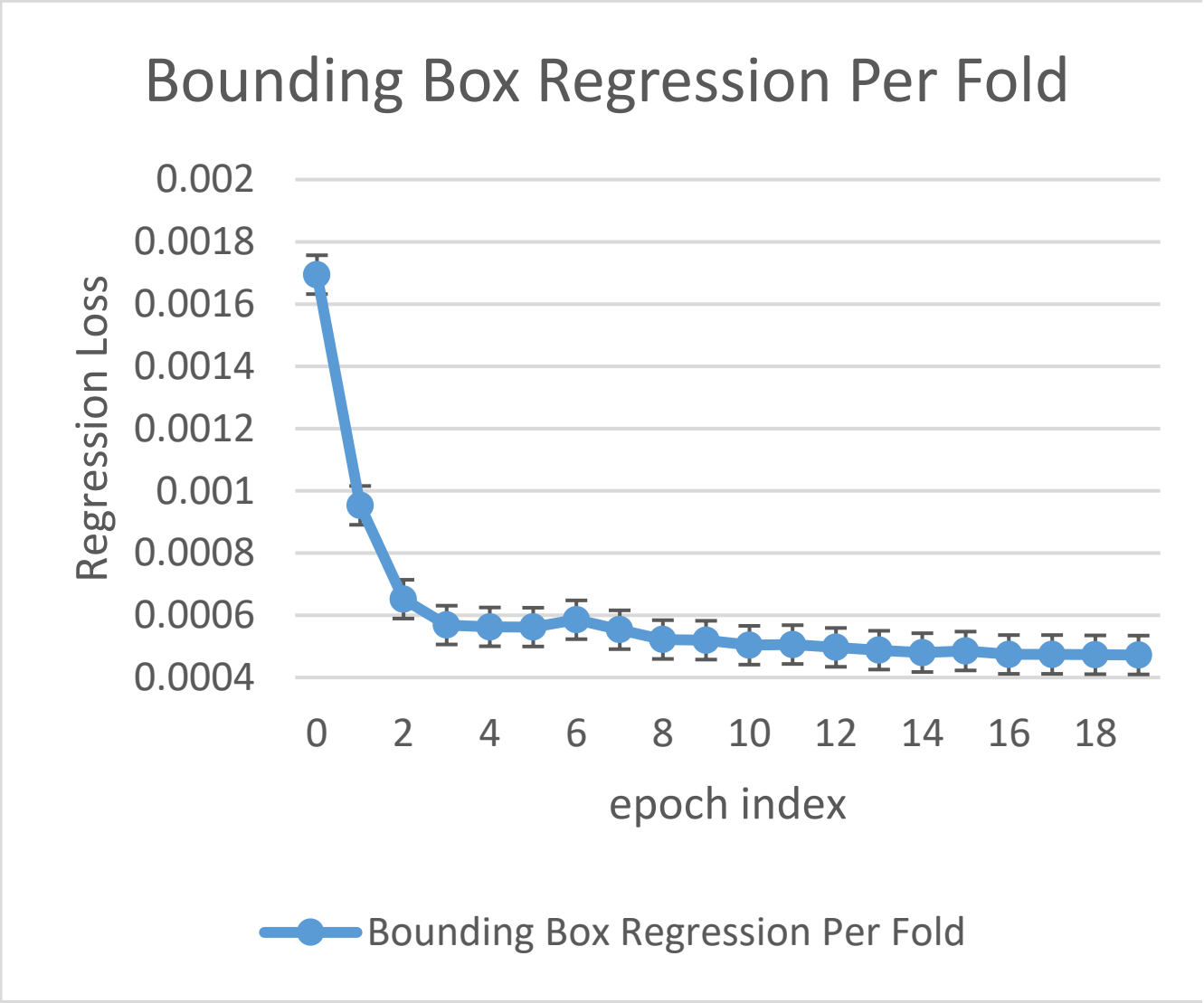

Fig. 6. Regression Loss graph (convergence) for the GMTI target localization using YOLO.

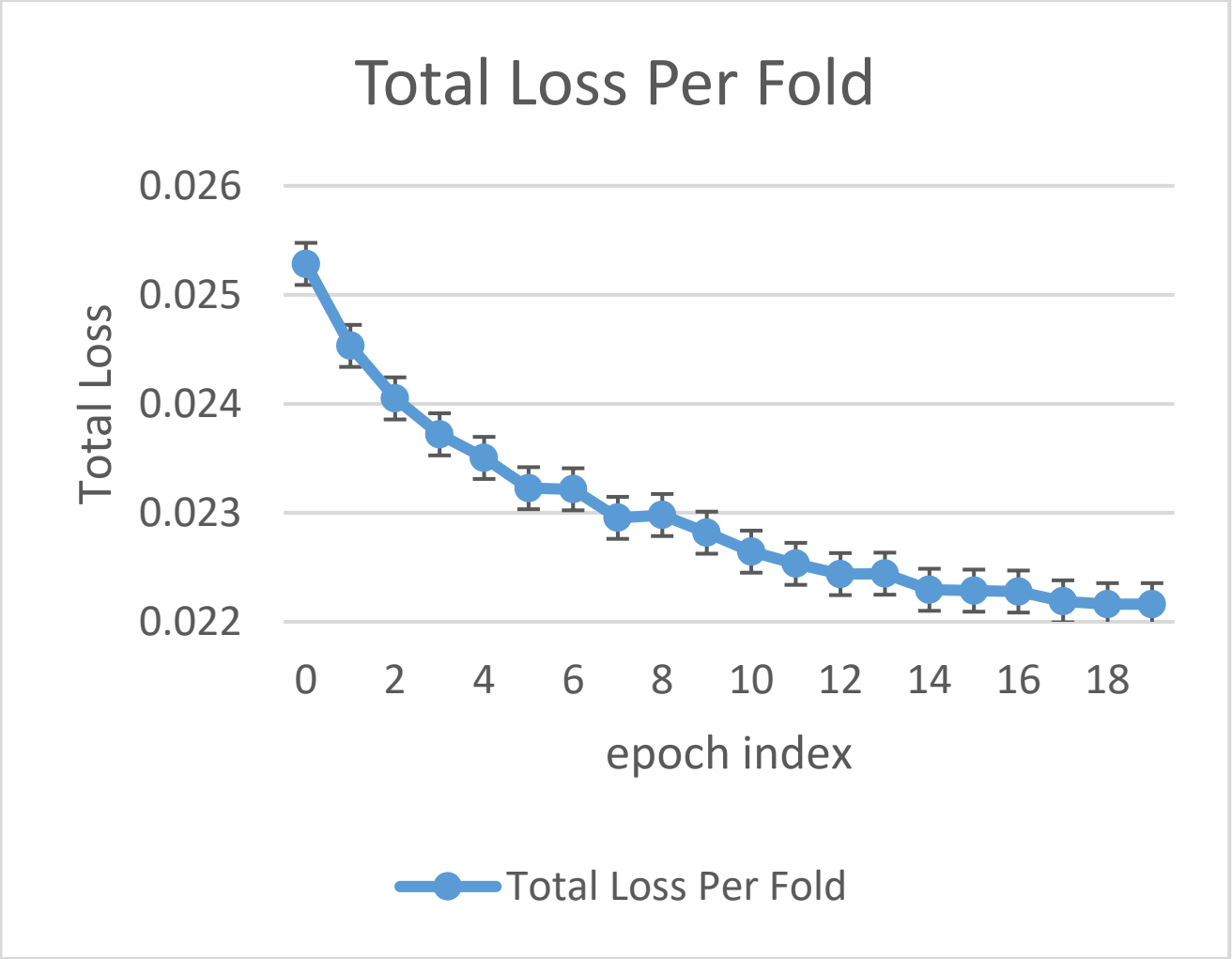

Fig. 7. Total Loss graph (convergence) for the GMTI target localization using YOLO.

First, we computed the MAE for all three networks. These results were compared with a simple ArgMax algorithm which would estimate the target location based on the maximum absolute value pixel on the RD map. The MAE is expressed in terms of the number of bins/pixels away from the true target location. Each range bin represents 0.0162 nautical miles and each Doppler frequency bin represents 3.4375 Hz. While the ArgMax and MobileNet algorithms are always designed to output a single target location as output, the RetinaNet and YOLO can output a varied number of bounding boxes. Therefore, we also compute the false positive and false negative rates to ensure the reduction in MAE does not come at the cost of these rates. From the results in Figure 5, we observe that all three CNNs outperform the ArgMax approach. It is understandable because the CNNs do not just use the peak but instead train on all the surrounding patterns in the RD map to obtain enhanced performance. These MAE numbers were averaged across 5 folds. The network obtaining the best performance was the YOLOv7. At this baseline stage, the algorithm designer can modify the CNN architectures until the performance meets a prescribed acceptance level. Since YOLO offers the best performance, we plot the loss functions for YOLO as a function of the epoch number. Figures 6 and 7 also include the 1-sigma error bars. The total loss in Figure 7 also includes the false positives and false negatives during the training stage of the YOLO.

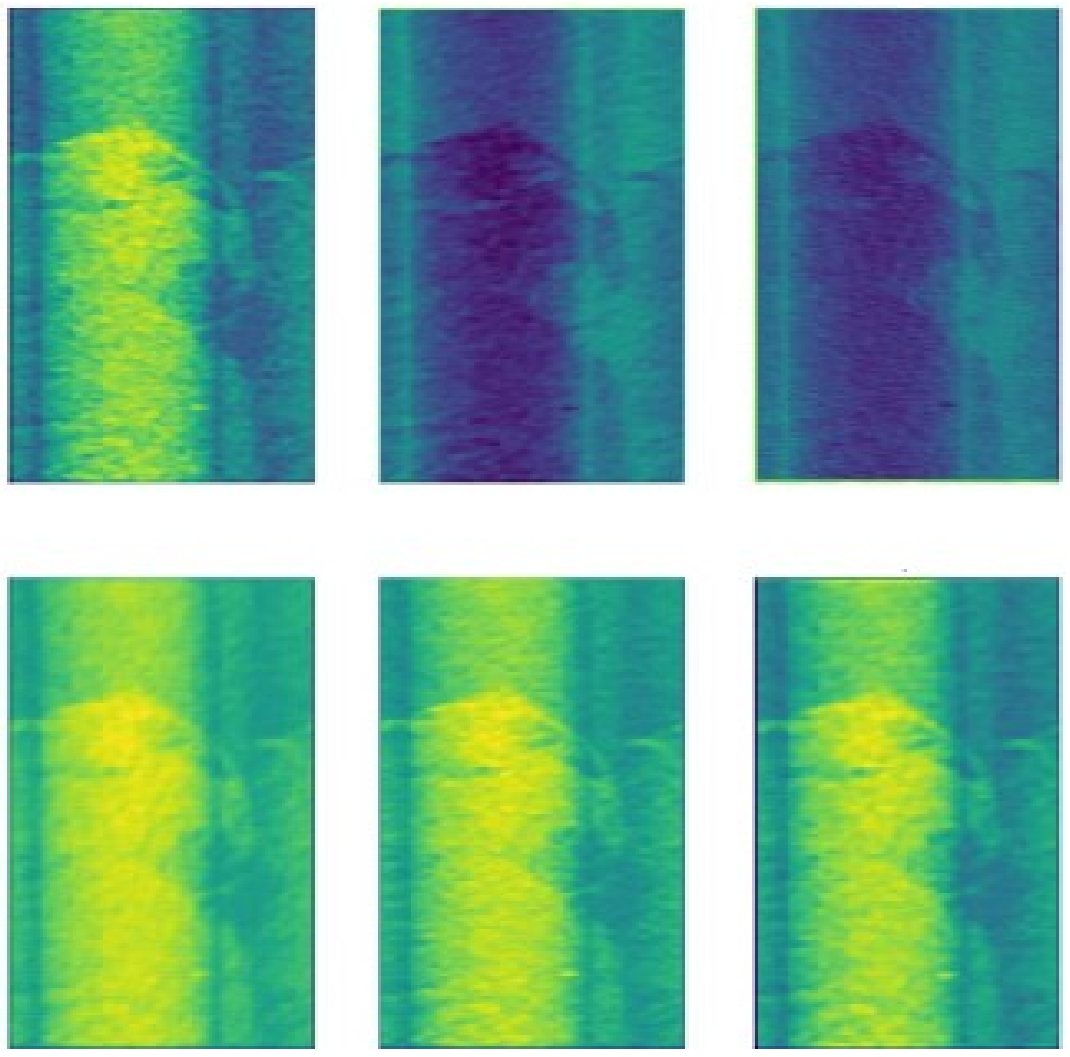

Fig. 8. Feature maps for GMTI target localization RetinaNet.

Recent advances in explainable AI (XAI) have developed tools for providing an enhanced understanding of the CNNs. One such tool is the convolutional feature map or activation map. They are a two-dimensional array or grid of numbers resulting from the application of convolutional filters (also called kernels) to an input image or a previous layer's feature map. In this case, the inputs are the RD maps. By studying these feature maps, we can identify areas in the images that indicate larger importance to final classification. In Figure 8, we plot some feature maps for RetinaNet. These feature maps show that the different filters extract areas of the input RD map that contain clutter and target returns and use that information to detect and identify the target of interest. This provides additional confirmation that the network has learned the desired features. Assuming the algorithms meet the prescribed acceptance levels on the baseline dataset, we move on to Phase II.

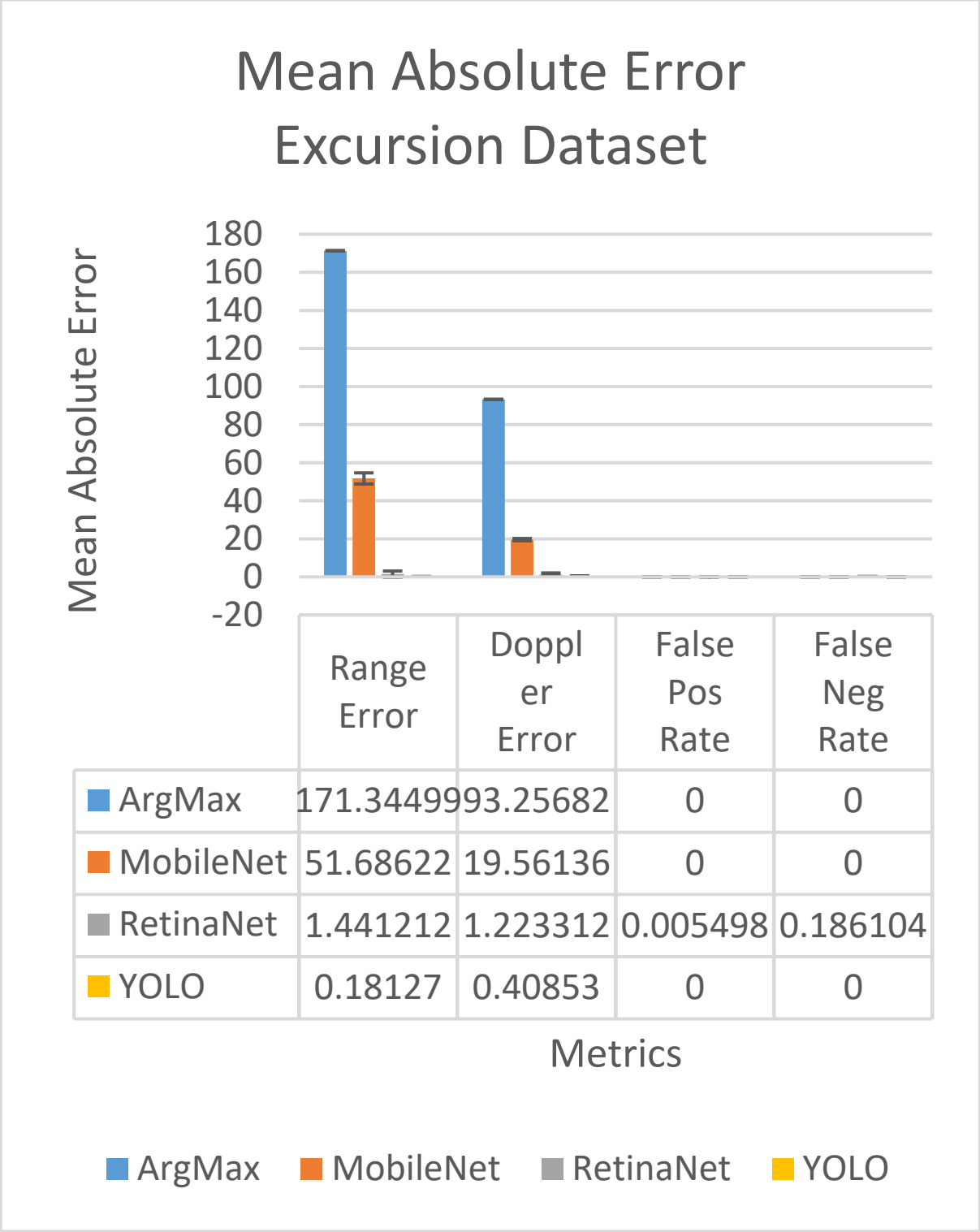

| | Range Error | Doppler Error | False Pos Rate | False Neg Rate |
|---|---|---|---|---|
| ArgMax | 171.34499 | 93.25682 | 0 | 0 |
| MobileNet | 51.68622 | 19.56136 | 0 | 0 |
| RetinaNet | 1.441212 | 1.223312 | 0.005498 | 0.186104 |
| YOLO | 0.18127 | 0.40853 | 0 | 0 |

Fig. 9. Range and Doppler MAE for all algorithms on the excursion dataset.

**Phase II**

Next, in Phase II, the performance of the same CNNs was analyzed on the excursion dataset. In Phase I, the digital twin for the environmental modeling was assumed to be perfect. However, in reality, no model is perfect. In this phase, we generate additional datasets that consist of stronger clutter returns (modeling errors) beyond what is expected to be encountered when deployed if the digital model was perfect. This can happen for a variety of phenomenological reasons such as heavy precipitation, etc. We generated this dataset within RFView by increasing the scattered power from each clutter patch by 6 dB. This excursion dataset also consisted of 5000 RD map samples. Testing the algorithms on this excursion dataset resulted in an increase in the mean localization error for all the algorithms (Figure 9), as expected.

At this stage, designers will determine whether the degraded performance is acceptable given the unlikeliness of the excursion from baseline modeling assumptions. In this case, the degradation is quite significant for some of the algorithms, so a re-design of some sort is likely warranted. Unfortunately, due to physics, there isn't significant structure to the clutter patterns that a CNN can use to improve target localization. This happens quite frequently in complex interconnected systems. However, the increased clutter power can be overcome by an increased antenna size which reduces the antenna beamwidth and thereby helps the target in the mainbeam stand-out against the background clutter returns. For simplicity, we increased the antenna size by a factor of 2 in both the dimensions. Therefore, the antenna now had 20 horizontal elements and 10 vertical elements. Since the system variables were changed, at this step, we needed to re-train the CNN before evaluating the target localization performance. We observe from Figure 10 that the re-trained CNN has gained back all the lost performance on the excursion dataset.

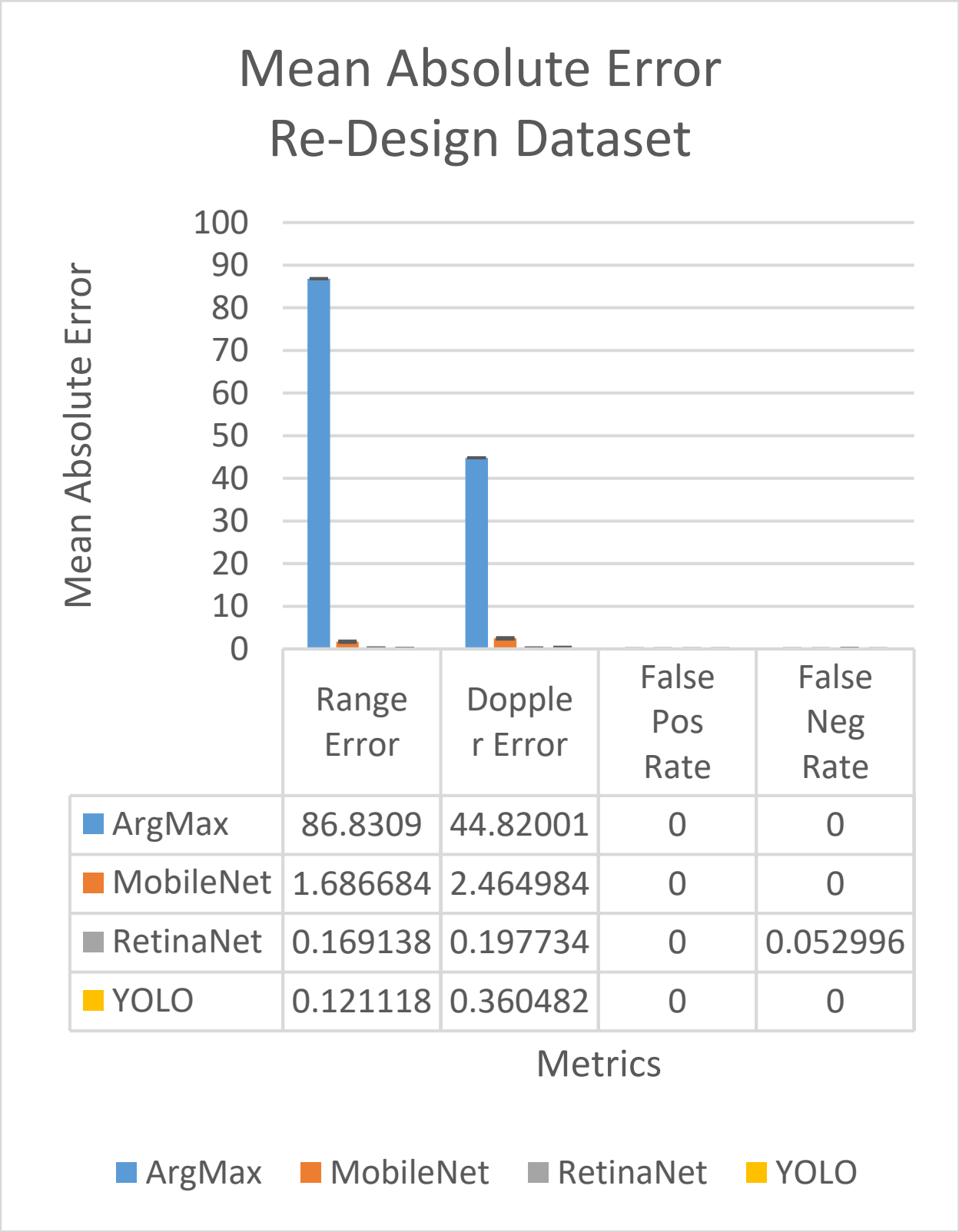

| | Range Error | Doppler Error | False Pos Rate | False Neg Rate |
|---|---|---|---|---|
| ArgMax | 86.8309 | 44.82001 | 0 | 0 |
| MobileNet | 1.686684 | 2.464984 | 0 | 0 |
| RetinaNet | 0.169138 | 0.197734 | 0 | 0.052996 |
| YOLO | 0.121118 | 0.360482 | 0 | 0 |

Fig. 10. Range and Doppler MAE for all algorithms on the re-design dataset with the larger antenna.

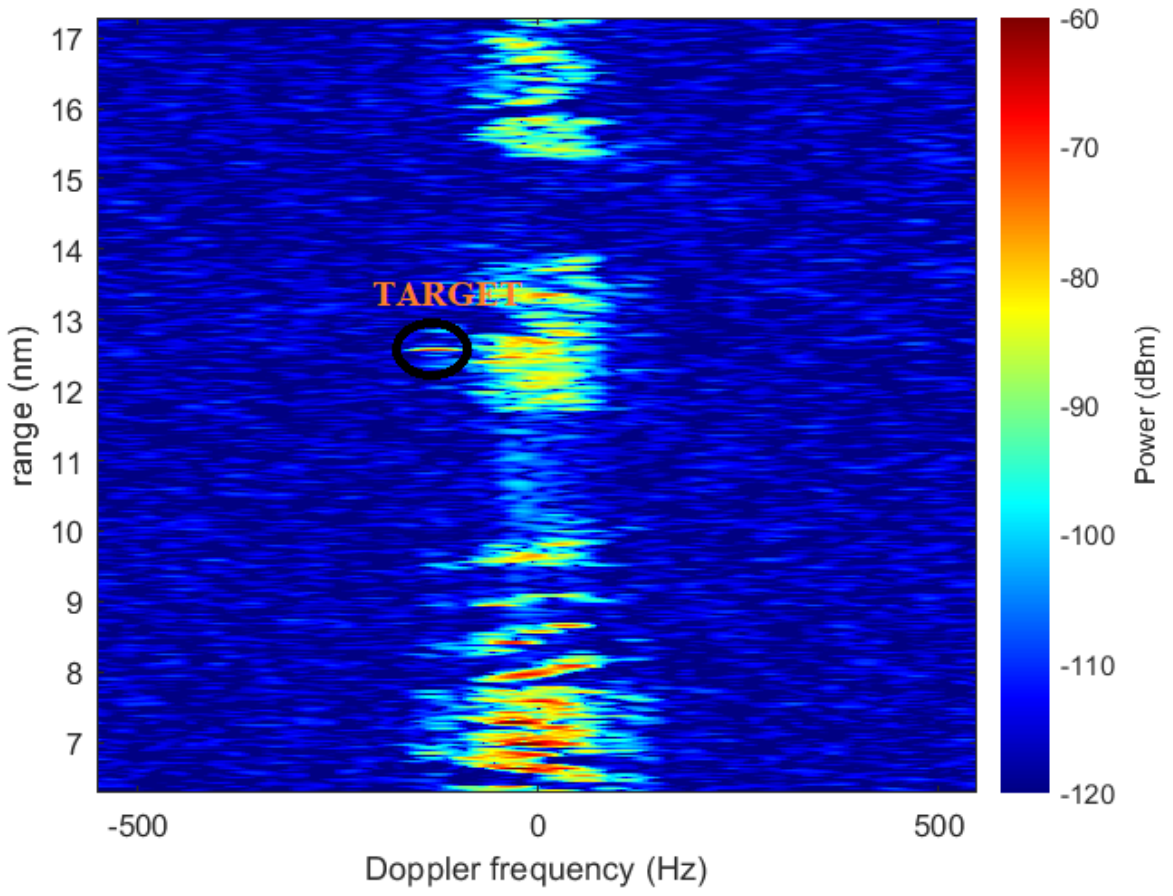

Fig. 11. Increased antenna size providing better separation between the clutter and target

The reason for the improvement in performance can be observed from Figure 11. When we have a larger antenna, that results in a smaller Doppler spread for clutter and thereby resulting in better separation between clutter and moving targets. In this example, Figure 11 is the same scenario as Figure 3, except for a larger antenna. Because of the narrower Doppler spread, in Figure 11, we are able to easily detect the target as opposed to Figure 3. So far, we have only used the MAE as the metric to analyze the algorithm performance. However, we believe the MAE does not capture the complete understanding of the CNN performance because a few bad estimates of the target location can completely degrade the average performance. Therefore, we also compute the percentage of samples where the estimate lies within 1 bin of the true location in both range and Doppler. We present these results for the best performing CNN YOLO in Figure 12. Across all folds, the estimate falls within one bin of the true location with a fairly high percentage of occurrence.

| Baseline Dataset | % Range | % Doppler |
|---|---|---|
| Fold 0 | 99.776 | 96.197 |
| Fold 1 | 99.692 | 90.769 |
| Fold 2 | 99.557 | 94.457 |
| Fold 3 | 100 | 95.595 |
| Fold 4 | 99.775 | 88.315 |

| Excursion Dataset | % Range | % Doppler |
|---|---|---|
| Fold 0 | 99.936 | 92.64 |
| Fold 1 | 99.97 | 90.312 |
| Fold 2 | 99.827 | 93.638 |
| Fold 3 | 99.808 | 93.47 |
| Fold 4 | 99.9 | 87.782 |

| Re-Design Dataset | % Range | % Doppler |
|---|---|---|
| Fold 0 | 99.884 | 93.605 |
| Fold 1 | 100 | 96.647 |
| Fold 2 | 99.651 | 93.83 |
| Fold 3 | 100 | 93.488 |
| Fold 4 | 100 | 94.941 |

Fig. 12. The percentage of samples where the YOLO target location estimate lies within 1 bin of the true location for the baseline, excursion, and re-design datasets.

At this stage of the new DE-T&E process, the system is deemed to have met acceptance testing specifications after the necessary system design modifications were made. However, as mentioned in Section 2, the new DE-T&E process also has a new "sustained testing" phase that leverages generative AI to search for so-called "Black Swan" events. This is discussed in the next section.

## V. PHASE III: HARNESSING GENERATIVE AI FOR "BLACK SWAN" EVENTS

In our example, we illustrate a method where the 'target detection/location system under test' is evaluated against RD inputs that obey valid statistics but were unanticipated (unknown - unknown) much as completely valid chess moves but unknown after years of study were revealed by Google's AlphaZero [15]. Generative AI (GAI) refers to deep learning methods used to predict or synthesize new results or data, based on what it has learned from extensive training sets. In the case of ChatGPT for example, it's used to synthesize plain text based on an input subject prompt [16]. Generative Adversarial Networks (GANs) are a type of neuromorphic architecture with two deep learning neural networks (DLNNs), a generator network and a discriminator network (see Figure 13). The generator tries to minimize the objective function while the discriminator tries to maximize it. The network is essentially a minmax game where the generator network tries to create images to fool the discriminator and the discriminator tries to detect synthetic images. Once the network is trained, the discriminator is removed and the generator operates alone. A common application for GANs is to amplify a limited number of input training examples with realistic synthetic images.

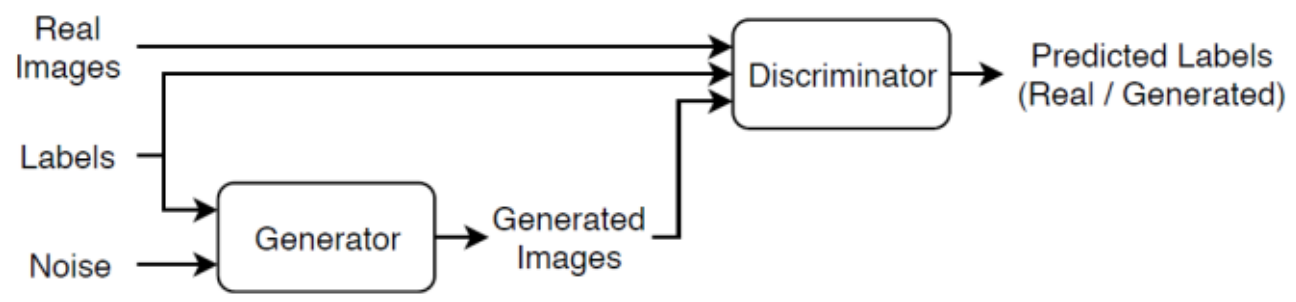

Fig. 13. Basic conditional GAN block diagram [17]

Data starved scenarios significantly benefit from this ability of a GAN to quickly generate large quantities of synthetic data from limited high-fidelity data samples. In the context of this paper, we need a mechanism to generate data representing "unknown unknowns". This goes beyond simple perturbations to the measurement models used within the "Digital Twin". These are events which are typically not predictable by humans and lead to data that the AI algorithms have never seen before. For the radar example presented in this paper, these events could be in the form of unknown scattering objects or swarm of objects appearing in the RD maps.

To demonstrate this concept of using GANs to generate RF data, we first trained a GAN using RFView generated radar clutter data. At the end of the training, the GAN, which is excited by noise signals can generate realistic looking clutter maps when provided with digital terrain elevation data (DTED) and Land Cover Land Use (LCLU) data as inputs. The key to completing this task is the availability of true clutter maps using RFView. These initial clutter maps aid in the training of the GAN which was then used to generate new synthetic clutter maps.

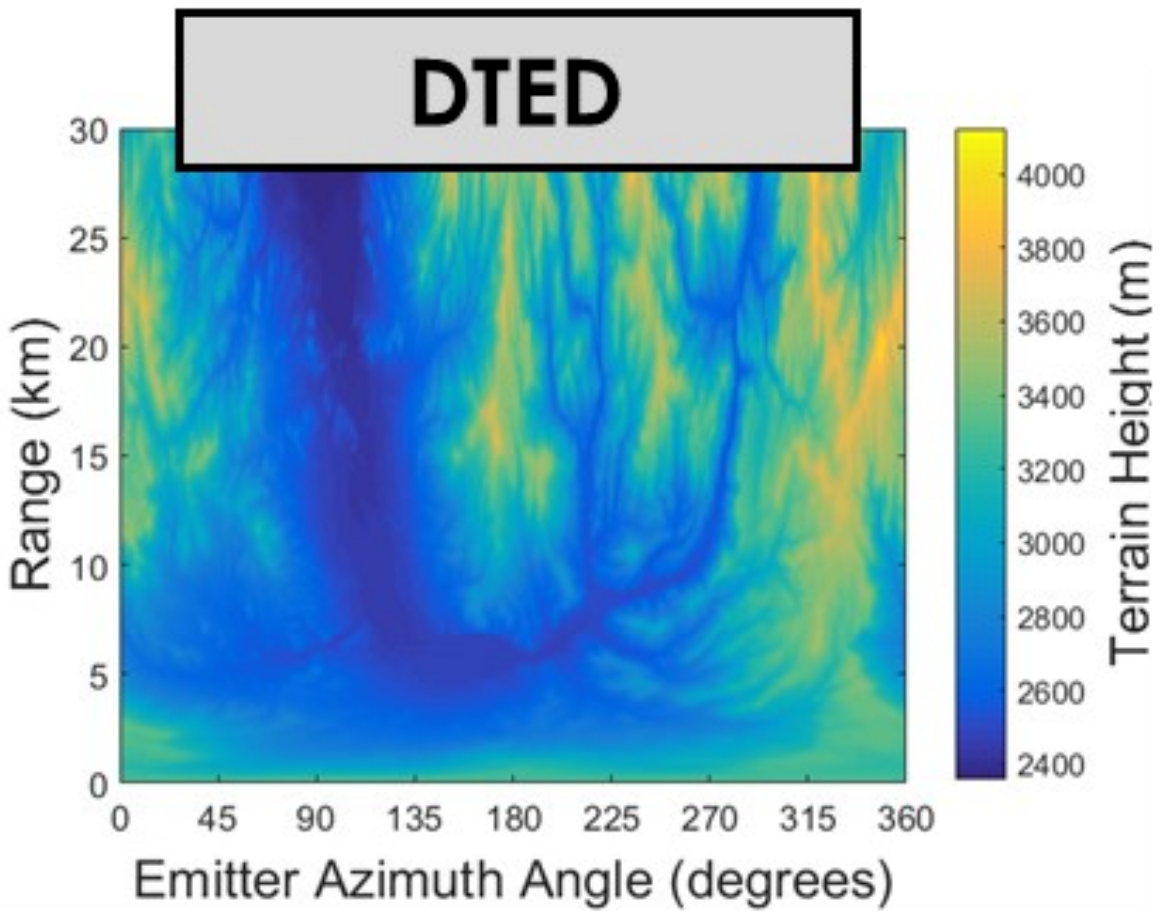

(a)

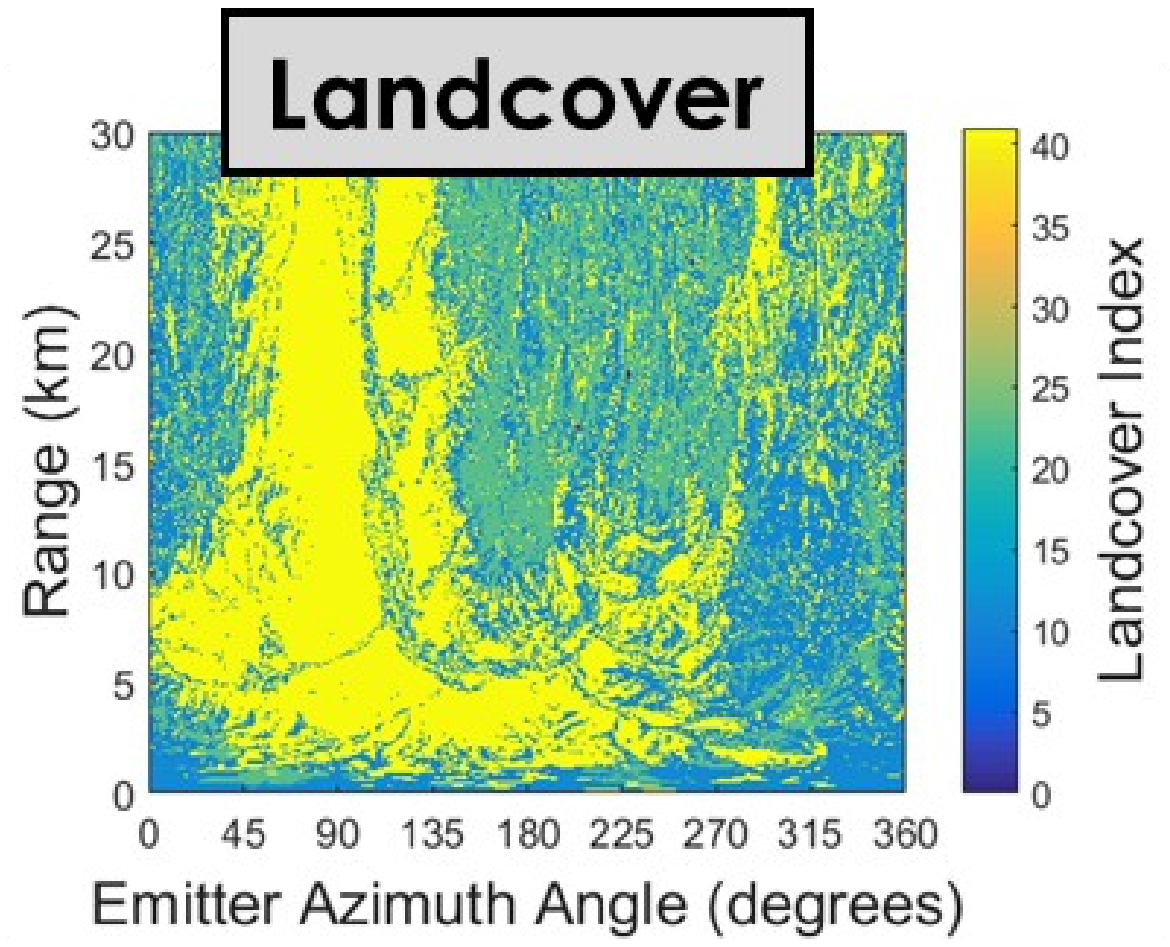

(b)

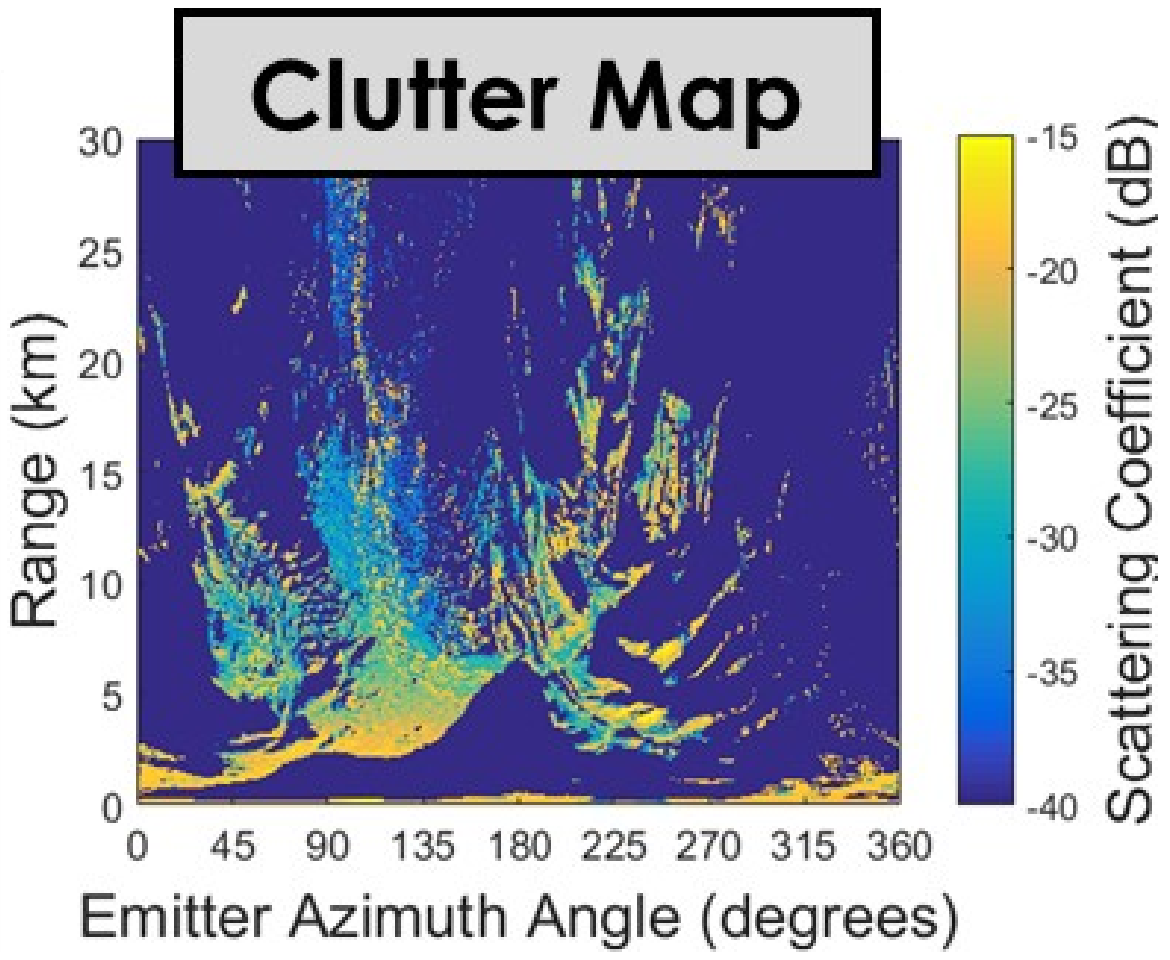

(c)

Fig. 14. (a) DTED input image, (b) landcover input image, (c) Output clutter map created using RFView.

For RF imagery, we chose to use the Isola GAN [18]. This GAN is good for modifying images where the resulting pixels are transformed through a non-linear operation but still localized with respect to the input training images. In our RF application, we convert DTED and landcover "images" into clutter maps. The features in the output images (such as landcover changes) are in approximately the same position as the input images. The shadows will be shifted and extended from the original features but are still generally localized with respect to the corresponding elevation changes (hills, valleys, etc.) in the DTED data. Thus, the Isola GAN is a good choice for image translation with localized feature transformations.

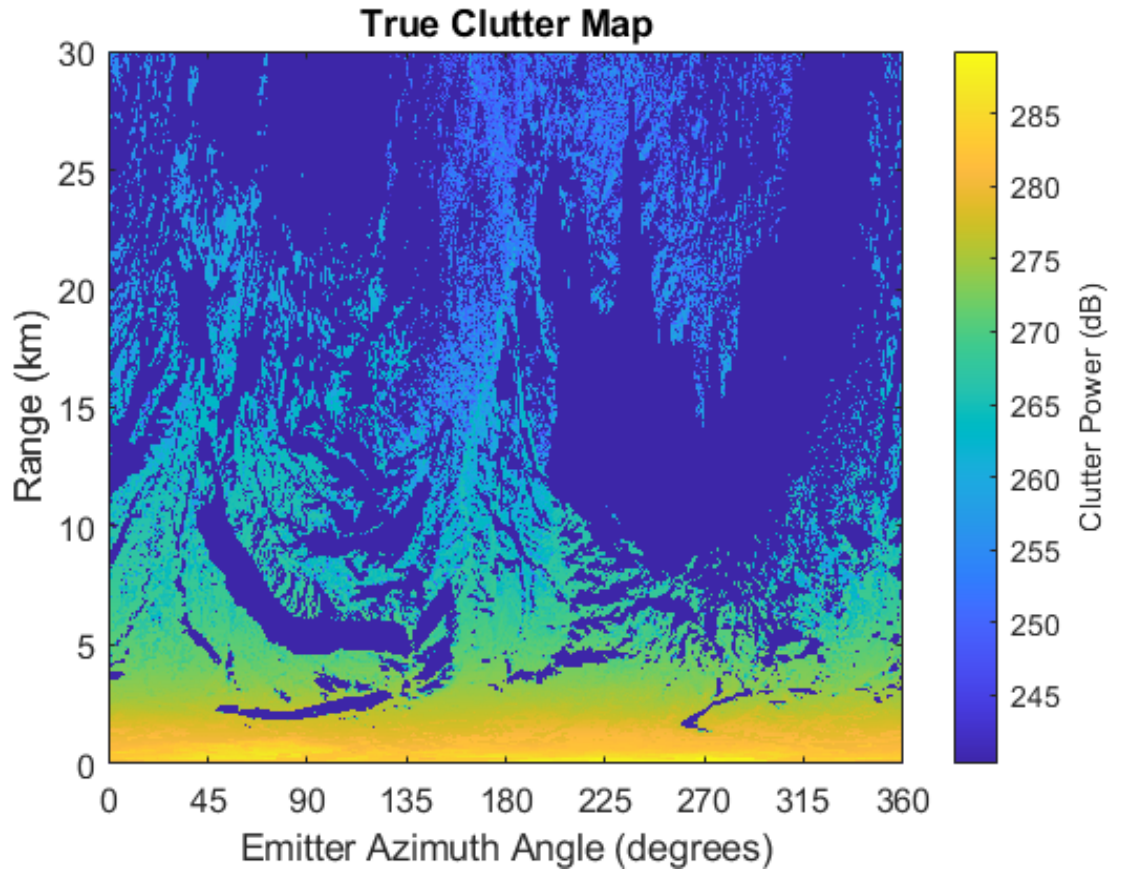

(a)

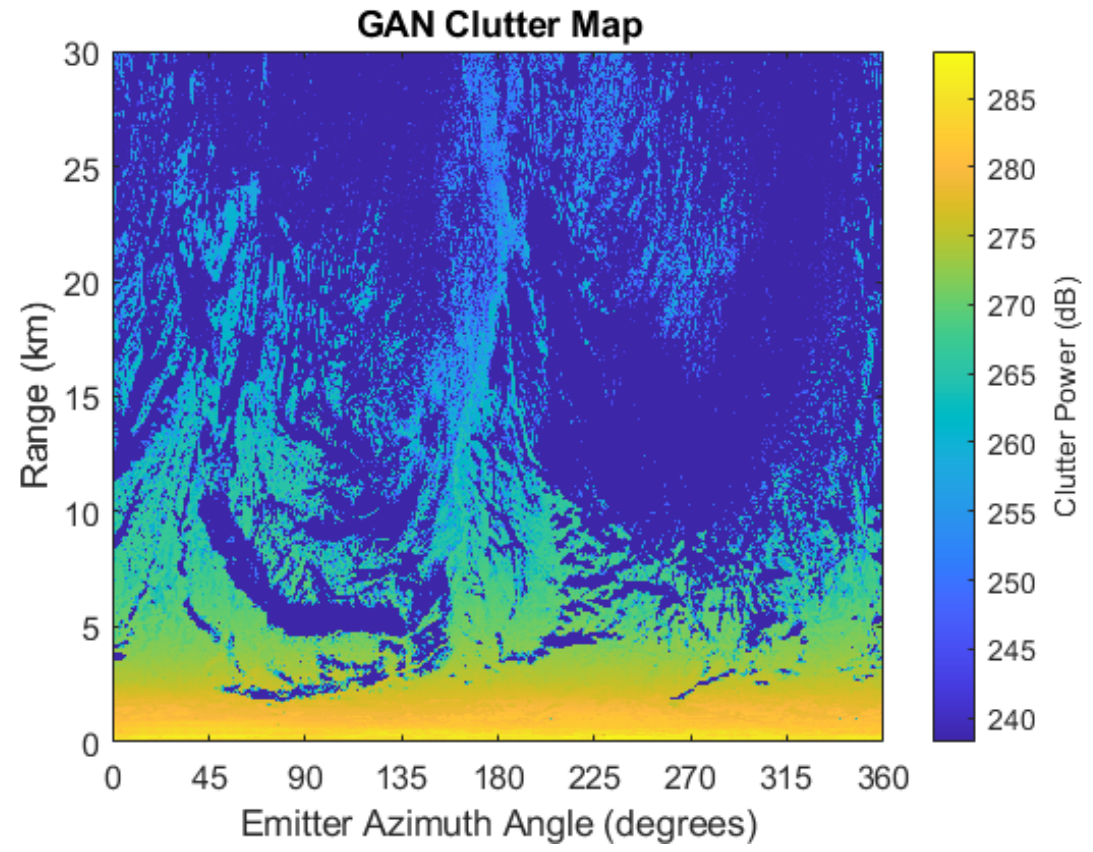

(b)

Fig. 15. (a) Example clutter map created using RFView and (b) Synthetic clutter map created using a GAN.

For each radar clutter image generated using RFView, the radar platform was moved to random locations and bearings in the CONUS. The altitude was fixed at 1 km above the average terrain height. The antenna was assumed to be omnidirectional. At each location, we calculated the clutter power versus range and azimuth as shown in Figure 14 using RFView. The goal is to train the GAN to generate similar clutter maps. Conditioning is very important for the radar training images. For this GAN application, DTED, landcover, and clutter images were scaled so that the pixel values ranged from 0 to 1. The log of the clutter values was used. The clutter power was compensated (by range squared) to mitigate the fall-off with range. In general, we will need to deconvolve the transmit waveform (if not an impulse).

The input data consisted of 1000 DTED, landcover images as input and 1000 clutter maps (created using RFView) as the output. A total of 600 input/output data sets were used for training and 400 were held back for validation. Training required approximately 6 hours. An example of the output clutter maps from the validation data set are shown in Figure 15 and Figure 16. The synthetic clutter maps closely resemble the output from conventional M&S. Rivers and lakes are also accurately predicted. A GAN is able to predict the output power accurately. The GAN is performing a complex non-linear conversion that utilizes landcover changes, shadowing, and scattering phenomenology from each landcover type. Most importantly, the image translation is occurring with extremely low timing latency (less than 4 ms) so that the ground clutter could be predicted in real time. The original RFView clutter maps required 900 ms per image. This demonstrates the capability of a GAN to generate realistic RF data much faster than traditional M&S.

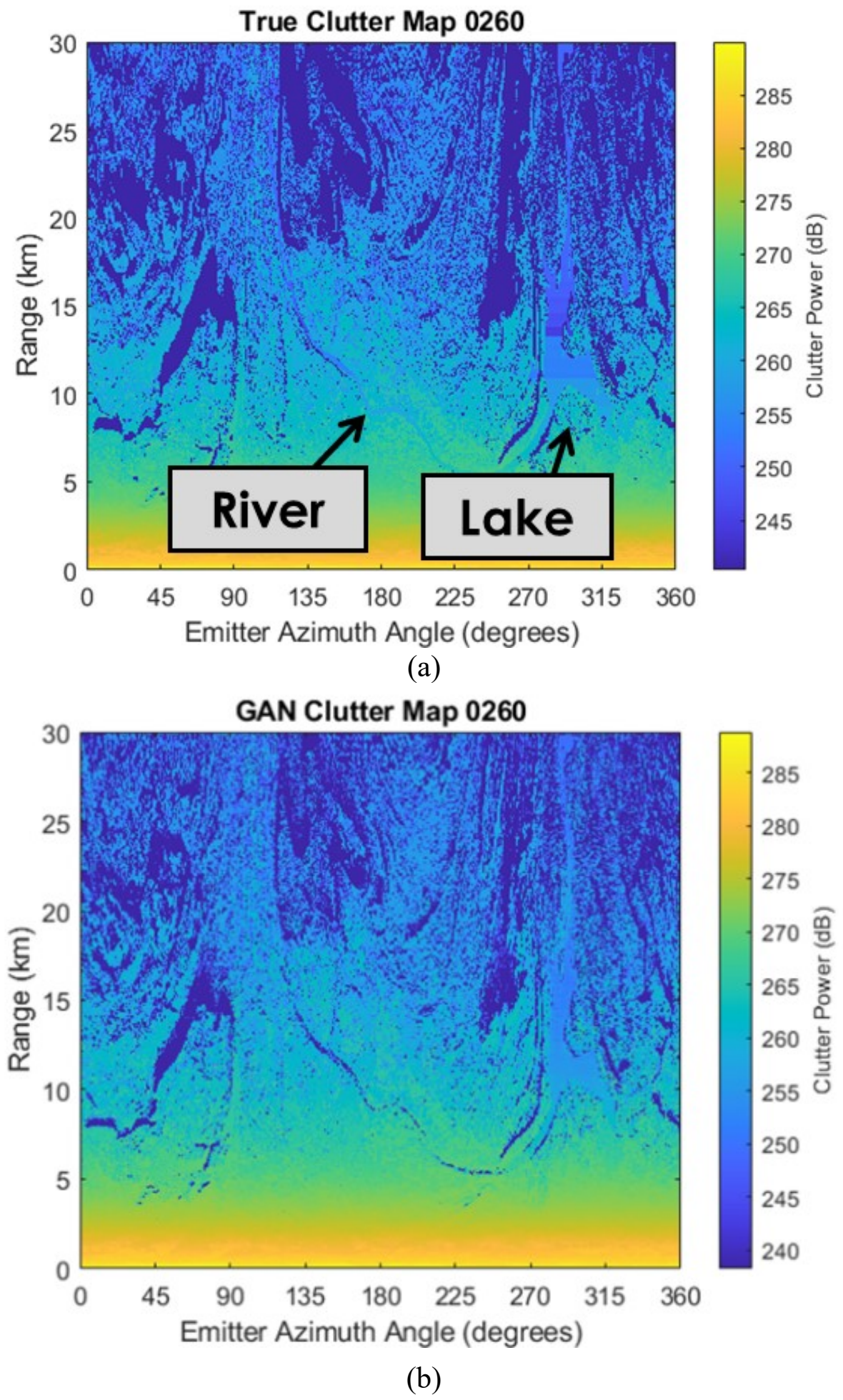

Fig. 16. (a) Example clutter map created using RFView and (b) Synthetic clutter map created using a GAN.

Along with the ability to facilitate the rapid generation of RF data, GANs will also serve as the mechanism to generate the Black Swan events in the context of this paper. As seen in Figure 13, the GAN data generation is excited by noise input. As long as the distribution of the noise random process used during training process of the GAN and the process used during the synthetic data generation follow the same statistical distribution, it is reasonable to assume that the data faithfully represents the characteristics of the data used for the training process. By introducing completely random variations to this distribution, one can make the GAN generate outputs that are not representative of the "known" events and this will create data with "unknown unknowns" that can be used to complete the three phased T&E process presented here. This GAN is envisioned to run concurrently with deployment. If a significantly deleterious event is discovered based on the synthetic data generated by the GAN, designers can decide what if any corrective action needs to be taken proactively. Of course, the radar example presented here is for simple illustrative purposes on how GANs can be trained to generate RF data. The ultimate goal of this process is to apply it to far more complex integrated systems and systems of systems thus not only addressing the current lack of testing approaches for but simultaneously implementing DoD Instruction 5000.97.

## VI. SUMMARY

In this paper, we introduced a Digital Engineering (DE) approach to T&E (DE-T&E) that can achieve requisite mil-spec statistical validation and simultaneously implement DoD Instruction 5000.97 [8]. An entirely new Phase was introduced to harness the power of AI to uncover so-called "Black Swan" events that might otherwise not be uncovered. It is intended to run concurrent with deployment. An illustration of these concepts was presented for a modern radar example employing deep learning AI.